\documentclass[letter, 11pt]{article}
\usepackage{adjustbox,amsmath,amstext,graphicx,amsfonts,amssymb,mathrsfs,amsthm,amsopn,geometry,graphics,rotating,longtable,lscape,multicol,pgfpages,xcolor,soul}
\usepackage[sectionbib]{natbib}
\usepackage[hyphens]{url}
\usepackage[hidelinks]{hyperref}
\usepackage{float,booktabs,array,caption,subcaption}
\usepackage{arydshln}   
\usepackage{booktabs}   
\usepackage{rotating}   
\usepackage{float}      
\hypersetup{breaklinks=true}

\definecolor{myblue}{RGB}{38,28,160}

\title{\textbf{The Role of Data and Metrics in Measuring Inequality Worldwide}\\[0.5em]
    \large\textit{A Tribute to Giovanni Andrea Cornia's Lifelong Work on the World Ginis}}

\author{Lidia Ceriani\thanks{lidia.ceriani@univr.it}\\[0.3em] University of Verona \and Paolo Verme\thanks{Corresponding author: paolo.verme@unibo.it}\\[0.3em]
\small University of Bologna}

\date{}

\begin{document}
\thispagestyle{empty}
\maketitle

\begin{abstract}
This paper pays tribute to Professor Giovanni Andrea Cornia's lifelong contributions to the measurement of global inequality. We review twelve world and regional databases of the Gini coefficient, illustrate their coverage, overlapping, and data gaps, and analyse the major sources of discrepancy among published Ginis. Merging all databases into a unified collection of over 122,000 observations spanning 222 countries from 1867 to 2024, we document how differences in welfare metrics, reference units, sub-metric definitions, post-survey adjustments, and survey design produce Gini estimates that diverge considerably---sometimes by as much as 50 percentage points---for the same country and year. We quantify pairwise cross-database discordance, document the income--consumption Gini gap by region and income group, and discuss the contributions of welfare metric and equivalence scale choices to cross-database dispersion. We extend the analysis with a dedicated discussion of comparability across time and across measurement dimensions, showing how multiple layers of methodological choice interact to make any single Gini figure a product of a complex chain of decisions that are rarely fully disclosed. Our analysis confirms that the choice of welfare metric remains the single most important source of cross-country non-comparability, while sub-metric definitions and equivalence scales introduce further systematic differences that are routinely overlooked in comparative work.
\end{abstract}

\noindent\textbf{JEL Codes:} D31, D63, I31, O15, C81 \\

\noindent\textbf{Keywords:} Gini coefficient; income inequality; consumption; welfare measurement; cross-country comparability; global databases; Giovanni Andrea Cornia

\newpage
\setcounter{page}{1}

\section{Introduction}
\label{sec:intro}

Giovanni Andrea Cornia (1947--2024) devoted much of his intellectual life to understanding how inequality shapes human development. From his pioneering work with UNICEF in the 1980s on the human costs of structural adjustment \citep{CorniaEtAl1987}, to his role in establishing and maintaining one of the first comprehensive world databases of inequality statistics, to his sustained scholarly engagement with the determinants and consequences of income inequality across the development spectrum, Cornia never ceased asking the fundamental question: what are the actual distributional consequences of economic policies and structural change?\footnote{This paper was originally prepared for the conference \textit{``Inequality and Development Revisited: The Legacy of Giovanni Andrea Cornia''}, organised in the summer of 2025 by the University of Florence in memory of Giovanni Andrea Cornia (1947--2024).}

Cornia's contribution to the measurement of global inequality is among the most enduring of his legacies. In the late 1990s and early 2000s, he played a central role at UNU-WIDER in building and expanding the World Income Inequality Database (WIID)---a project that grew out of the earlier Deininger--Squire dataset \citep{DeiningerSquire1996} and evolved into one of the most comprehensive collections of inequality statistics ever assembled. By providing researchers worldwide with a single point of access to Gini coefficients, income shares, and distributional data drawn from hundreds of household surveys and secondary sources, the WIID transformed empirical inequality research. It allowed scholars to move beyond country-specific studies and ask genuinely global questions about the trajectory of inequality, the relationship between growth and distribution, and the effects of trade liberalisation, fiscal policy, and technological change on income dispersion.

Cornia used this infrastructure, and the research it enabled, to advance a substantive agenda. His edited volume on inequality in the era of liberalisation \citep{Cornia2004} synthesised evidence from dozens of countries and documented a widespread tendency toward rising inequality during the 1980s and 1990s---a finding that challenged the optimistic predictions of trade-based models and highlighted the distributive costs of rapid opening. His work with Kiiski and Addison on post-WWII trends in income distribution \citep{CorniaKiiski2001, CorniaAddison2004} provided some of the most careful empirical documentation of the so-called ``inequality turn'' of the late twentieth century. And his abiding interest in Latin America---a region that experienced some of the most dramatic swings in inequality of the past half century---generated analysis that was simultaneously rigorous, policy-relevant, and historically informed.

This paper pays tribute to Professor Cornia's contributions by revisiting the world of Gini data that he helped create and taking stock of where the science of global inequality measurement stands today. Our exercise covers twelve world and regional databases of the Gini coefficient and pursues five objectives: (1)~reviewing the databases and their methodological characteristics; (2)~illustrating coverage, overlapping, and data gaps; (3)~understanding sources of discrepancy among published Ginis; (4)~quantifying the magnitude of these discrepancies; and (5)~generating a larger, consolidated collection of world Ginis that the research community can use as a starting point for future work.

This is no easy task because understanding the Gini figures published by different databases requires a level of detail that most databases do not provide. No database reviewed here supplies a complete list of items included in the welfare metric used to estimate the Gini. The quantity and quality of metadata varies widely across sources. This is therefore a necessarily incomplete exercise, but one that aims to improve on the current state of understanding of the world Ginis---a goal that Cornia would have recognised and endorsed.

The paper is organised as follows. Section~\ref{sec:databases} describes the twelve databases we cover and their key methodological characteristics. Section~\ref{sec:unified} describes the new unified database and provides stylised facts about global Gini coverage. Section~\ref{sec:discrepancy} analyses the main sources of discrepancy in depth. Section~\ref{sec:quantifying} quantifies cross-database discordance through a series of new empirical exercises. Section~\ref{sec:casestudy} uses Colombia as a detailed country illustration, and Section~\ref{sec:conclusion} concludes with reflections on Cornia's legacy.

\section{Global and Regional Collections of Gini Measures}
\label{sec:databases}

We cover twelve databases, distinguishing between global and regional databases, and between primary and secondary data sources. In this context, primary sources refer to databases that compute inequality indicators—such as Gini coefficients—directly from the underlying microdata (e.g., household surveys or administrative records). Secondary sources, by contrast, are collections of inequality indicators compiled from other studies, reports, or databases, rather than calculated directly from microdata by the data provider. The complete list is provided below; A detailed methodological overview of each is available upon request.

\bigskip
\noindent\textbf{Global databases}

\begin{itemize}
    \item \textit{Primary sources}
    \begin{itemize}
        \item LIS --- Luxembourg Income Study (Inequality and Poverty Key Figures)
        \item World Bank --- Poverty and Inequality Platform (PIP)
        \item WID --- World Inequality Database (Paris School of Economics and the University of California, Berkeley)
    \end{itemize}
    \item \textit{Secondary sources}
    \begin{itemize}
        \item UNU-WIDER --- World Income Inequality Database (WIID)
        \item All the Ginis (ATG) --- Stone Center on Socio-Economic Inequality (maintained by Branko Milanovic)
        \item SWIID --- Standardised World Income Inequality Database (maintained by Frederick Solt)
    \end{itemize}
\end{itemize}

\noindent\textbf{Regional databases}

\begin{itemize}
    \item \textit{Primary sources}
    \begin{itemize}
        \item CEPAL/ECLAC --- Economic Commission for Latin America and the Caribbean
        \item CEDLAS --- Center of Distributive, Labor and Social Studies (SEDLAC database)
        \item IDB --- Inter-American Development Bank (Soci\'{o}metro)
        \item Eurostat --- Income and Living Conditions (EU-SILC)
    \end{itemize}
    \item \textit{Secondary sources}
    \begin{itemize}
        \item OECD --- Income Distribution Database (IDD)
        \item ADB --- Asian Development Bank (Key Indicators Database)
        \item Afristat --- Conditions de vie des m\'{e}nages
    \end{itemize}
\end{itemize}

These twelve databases constitute the largest and most widely used collections of Ginis worldwide. Table~\ref{tab:databases} provides a comparative overview of their key methodological characteristics.

\medskip
\begin{center}\textit{[Table 1 about here]}\end{center}
\medskip

 The databases differ along several dimensions that directly affect the Gini values they report: the temporal and geographic coverage,  the welfare concept used (income versus consumption, and gross versus net income), the level of disaggregation (overall population, urban/rural, gender, age, and other), the level of aggregation (household, tax unit, individuals), the reference unit (per capita, adult equivalent, or household), and the degree of reliance on secondary sources.

\section{A Unified Database}
\label{sec:unified}

As a first exercise, we merged the twelve databases into one, generating the largest database of world Ginis to date. The resulting database includes 122,351 observations and spans 222 countries and territories from 1867 to 2024 (Table~\ref{table:sumstats}). The period with the best coverage runs from roughly 2003 to 2017, which all databases cover without exception.

\medskip
\begin{center}\textit{[Table 2 about here]}\end{center}
\medskip
 The early post-World War II period is covered by only a few databases, and the years prior to 1900 by a single source---the WID, which reconstructs historical inequality using a variety of archival sources.

The variation in temporal coverage reflects the history of household survey programmes worldwide. Income and consumption surveys emerged after World War II in high-income countries and expanded significantly in middle- and low-income countries starting from the 1980s, partly in response to the World Bank's Living Standards Measurement Study programme and related initiatives. Results for the most recent years appear in world databases with some delay, owing to the time needed to release microdata and the irregular update schedules of some sources.

By merging these databases, we are also able to construct a series of harmonised variables capturing the welfare metric (income, consumption, or expenditure), the type of metric (gross, net, or mixed income), the reference unit (per capita, adult equivalent, household), the data source, the survey design characteristics, and many other features. One evident limitation is that these features are fragmentary across observations. Aside from the income--versus--consumption distinction---which is generally reported---only some databases include the kind of methodological detail needed to fully characterise the welfare aggregate. It is, for example, very difficult to obtain complete information on the sub-items included in income aggregates, such as government transfers and in-kind benefits, and almost impossible to know the details of imputation procedures for items such as imputed rent or home production. The absence of this information makes meaningful comparison across Ginis arduous, but merging databases at least allows us to identify precisely where the gaps lie.

\begin{figure}[H]
{\centering \includegraphics[width=1\linewidth,height=\textheight,keepaspectratio]{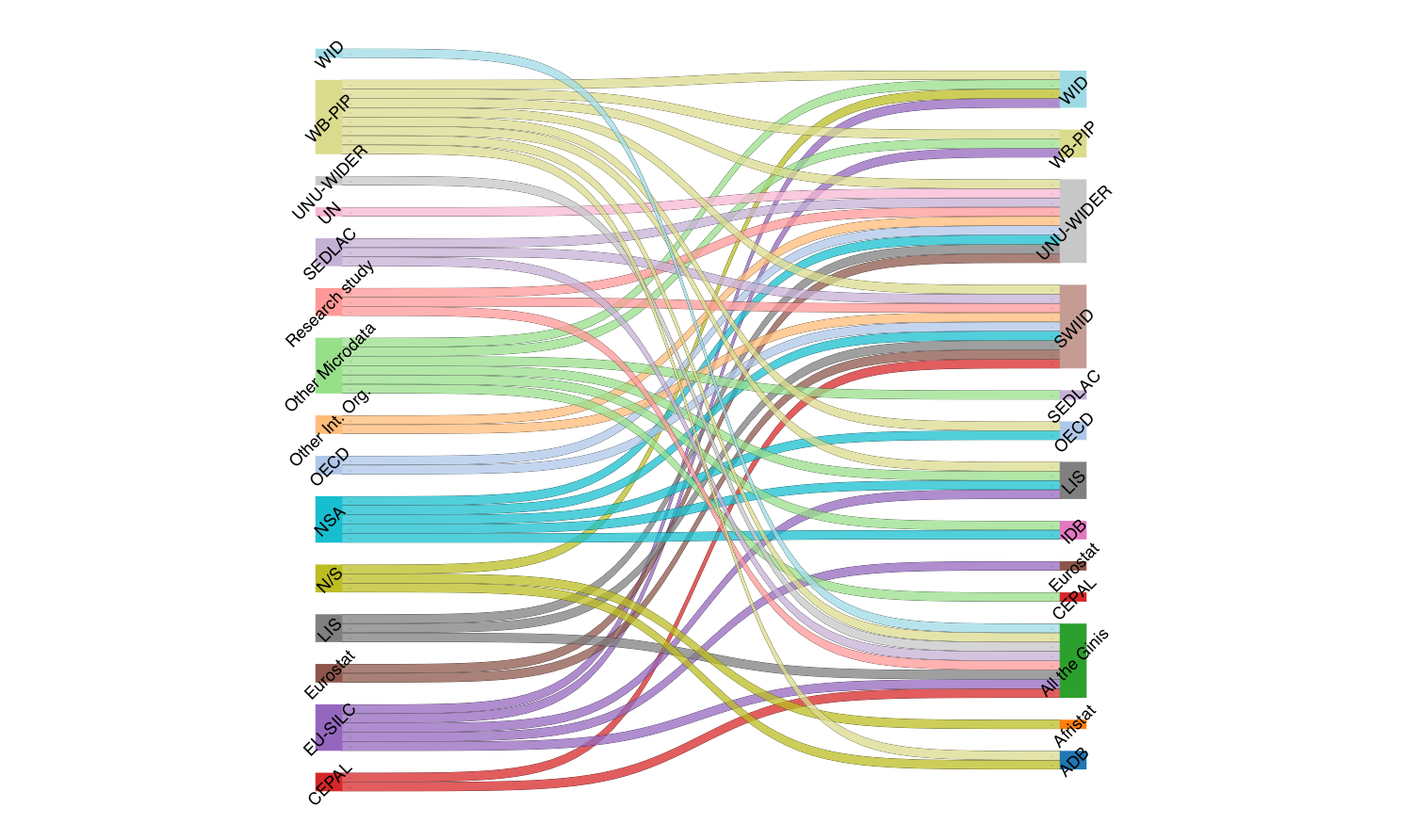}}
\caption{Data Sources and Destinations (Sankey Diagram)}
\label{fig:sources}
\end{figure}

Merging databases also allows us to map data provenance across sources. Figure~\ref{fig:sources} presents a Sankey diagram of these genealogical relations.

\medskip
\begin{center}\textit{[Figure 1 about here]}\end{center}
\medskip

 On the left-hand side are the data origins (primary surveys, national statistical agencies, and existing secondary databases) and on the right-hand side are the destinations (the twelve databases covered here). The global secondary databases---UNU-WIDER WIID, SWIID, and All the Ginis---draw on the largest number of sources. These sources overlap substantially, making the secondary databases very similar to one another; but the overlaps are not complete, which keeps them distinct. National Statistical Agencies (NSAs) feed into the largest number of destination databases, since they provide the household surveys that are the ultimate source of virtually all Ginis.

\section{Sources of Discrepancy}
\label{sec:discrepancy}

Measuring economic inequality matters because it reveals how countries distribute resources through the system of production---via capital, labour, and taxes---and redistribute resources through public spending financed by taxes. The Gini coefficient is, arguably, the most widely used summary statistic of inequality. Yet the science of estimating it is fraught with complications and technicalities that make comparison across countries and time periods genuinely challenging.

It is not uncommon to find estimates of inequality for the same measure, country, and year that diverge substantially. \citet{Verme2014}, for example, report multiple Gini estimates for Egypt based on the same underlying data source. What explains these differences? In principle, any aspect of survey design, data processing, or methodological choice that affects the shape of the income or consumption distribution can produce divergent results. This section focuses on the principal sources of discrepancy, which we group into six categories: survey design; survey implementation; welfare metrics; welfare sub-metrics; reference units; and post-survey adjustments.

\subsection{Survey Design}

Even in the absence of technical errors, survey design involves normative choices that affect any distributional statistic. One important example is the inclusion or exclusion of marginalised population groups: prisoners, residents of institutions, homeless people, refugees, and internally displaced persons (IDPs). For most countries, these groups represent less than one percent of the population and their exclusion has only modest effects on the Gini. But in countries that host large refugee populations---Lebanon after the Syrian crisis being the most dramatic recent case, where refugees have at times constituted more than 20 percent of the resident population---the treatment of these groups can change the measured level of inequality substantially (\citealp{CerianiVerme_2022}).

A second normative choice concerns the reference period used for income or consumption measurement, particularly when measurement relies on recall questions. Short reference periods reduce recall error but increase the volatility of measured income; long reference periods smooth idiosyncratic shocks but may introduce systematic recall bias. These choices affect the quality of responses and the item non-response rate, and ultimately the degree of measurement error in the distribution.

Neither of these choices is routinely disclosed in the metadata accompanying world Gini databases, which means that end users comparing Ginis across sources may unknowingly be comparing estimates that differ for reasons of survey design rather than economic substance.

\subsection{Survey Implementation}

Measurement errors are a persistent feature of income and consumption surveys. Income and consumption are particularly prone to mismeasurement: income fluctuates, involves multiple sources that respondents may not report fully, and is subject to deliberate concealment; consumption is complex to measure accurately across diverse household types and requires careful attention to unit of measurement.

Of equal or greater importance is the problem of non-response, which may manifest as item non-response (missing values for particular income components) or unit non-response (missing households). If non-response is Missing Completely At Random (MCAR), the post-survey sample remains representative and any distributional statistic computed from it will be unbiased. But if non-responses are Missing At Random (MAR), or worse Missing Not At Random (MNAR)---as is well known to be the case for high incomes---then ignoring missing observations will bias inequality estimates downward. The concentration of non-responses at the upper tail of the income distribution means that even a small share of missing high-income households can substantially depress the estimated Gini. Very rarely are users of world Gini data informed about the presence of non-responses and how they have been treated.

\subsection{Welfare Metrics}

The most fundamental choice in measuring inequality is the selection of the welfare metric: income or consumption. These measures are structurally different---one captures income flows, the other actual consumption expenditure---and they differ in the degree of mismeasurement and non-response they suffer. In countries where work is largely informal and income is erratic, occasional, or difficult to capture, consumption may be the more reliable indicator of living standards. In countries with well-developed formal labour markets and reliable income records, income may be preferable.

\begin{figure}[H]
{\centering \includegraphics[width=1\linewidth,height=\textheight,keepaspectratio]{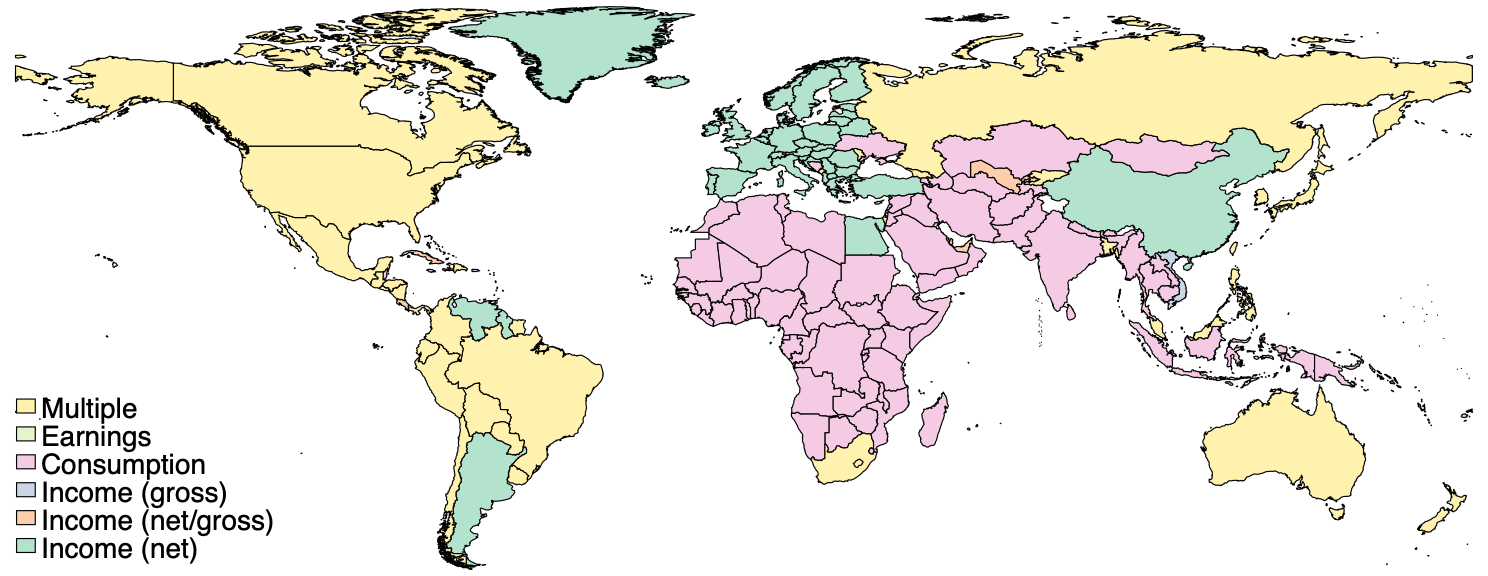}}
\caption{Countries' Prevalent Welfare Metric (most recent observation)}
\label{fig:maptype}
\end{figure}

Figure~\ref{fig:maptype} maps the world according to the welfare metric predominantly used in each country's most recent observation in our database. There is a clear geographical divide: low-income countries, particularly in Sub-Saharan Africa and South Asia, rely predominantly on consumption; middle- and high-income countries---across the Americas, Europe, and including China---use predominantly income or mixed metrics. This structural difference in data collection practice makes comparisons of Ginis between countries at different levels of development inherently problematic.

\medskip
\begin{center}\textit{[Figure 2 about here]}\end{center}
\medskip

\begin{figure}[H]
{\centering \includegraphics[width=1\linewidth,height=\textheight,keepaspectratio]{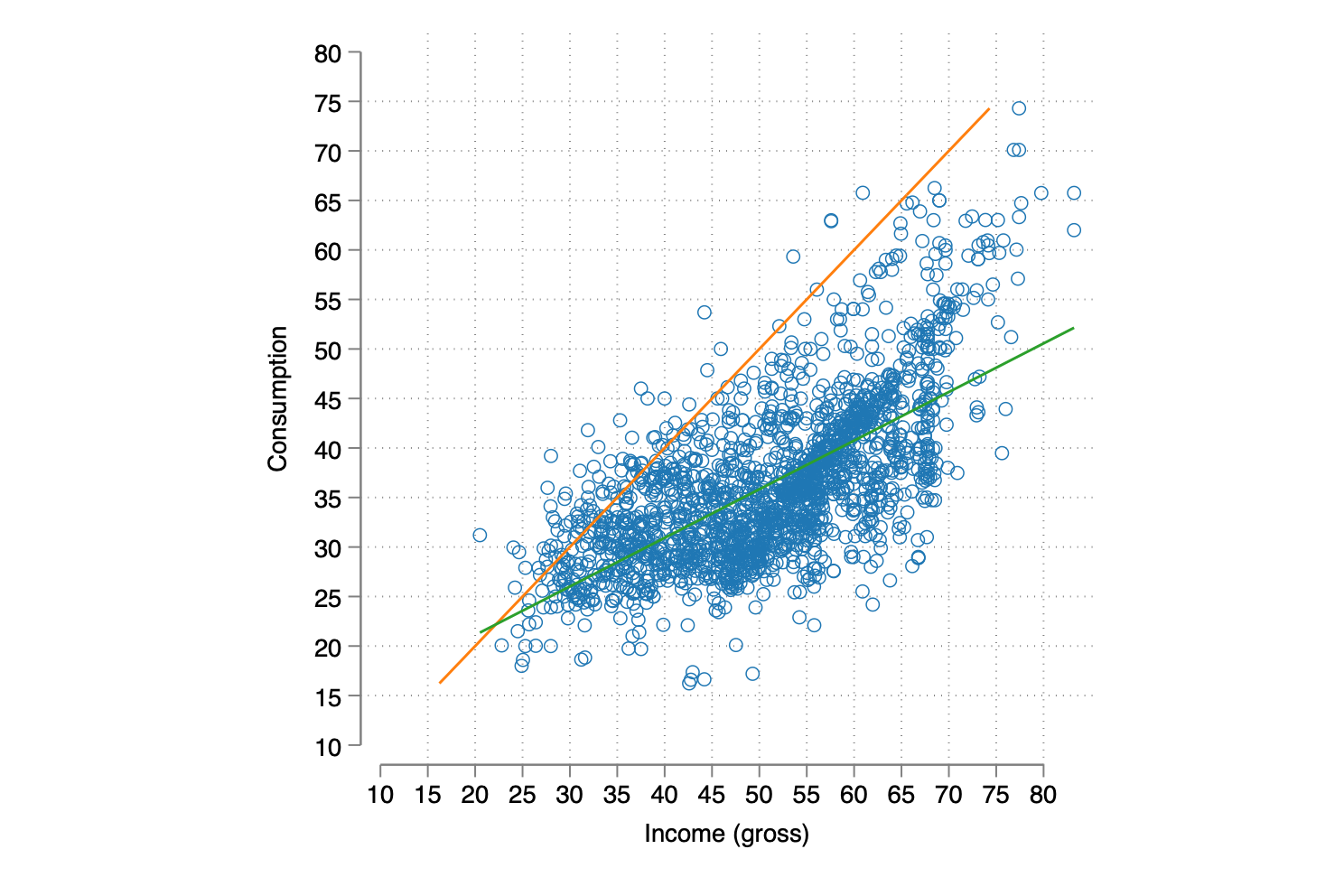}}
\caption{Ginis From Gross Income and Consumption (same country-year)}
\label{fig:incgross}
\end{figure}

Figure~\ref{fig:incgross} compares, for the same country-year observations, Gini estimates derived from income and from consumption.

\medskip
\begin{center}\textit{[Figure 3 about here]}\end{center}
\medskip

 The scatter of points deviates markedly from the 45-degree line of perfect concordance. Ginis derived from gross or net income are consistently and considerably larger than those derived from consumption---an expected result, since income is more dispersed than consumption, but the magnitude of the difference is striking. For some country-year observations, the gap between income and consumption Ginis reaches 50 percentage points. A consumption Gini of 30 can correspond to an income Gini of 70 or more for the same country in the same year. Researchers who combine income and consumption Ginis without explicit controls for this distinction risk severe comparability problems.

\subsection{Welfare Sub-metrics}

More subtle, and harder to detect from standard database metadata, are differences arising from the precise components included in the welfare aggregate. Consider income as the chosen metric: this may be measured including or excluding different components, such as government cash transfers, pensions, imputed rent, home production, or in-kind benefits from public services.

\begin{figure}[H]
{\centering \includegraphics[width=1\linewidth,height=\textheight,keepaspectratio]{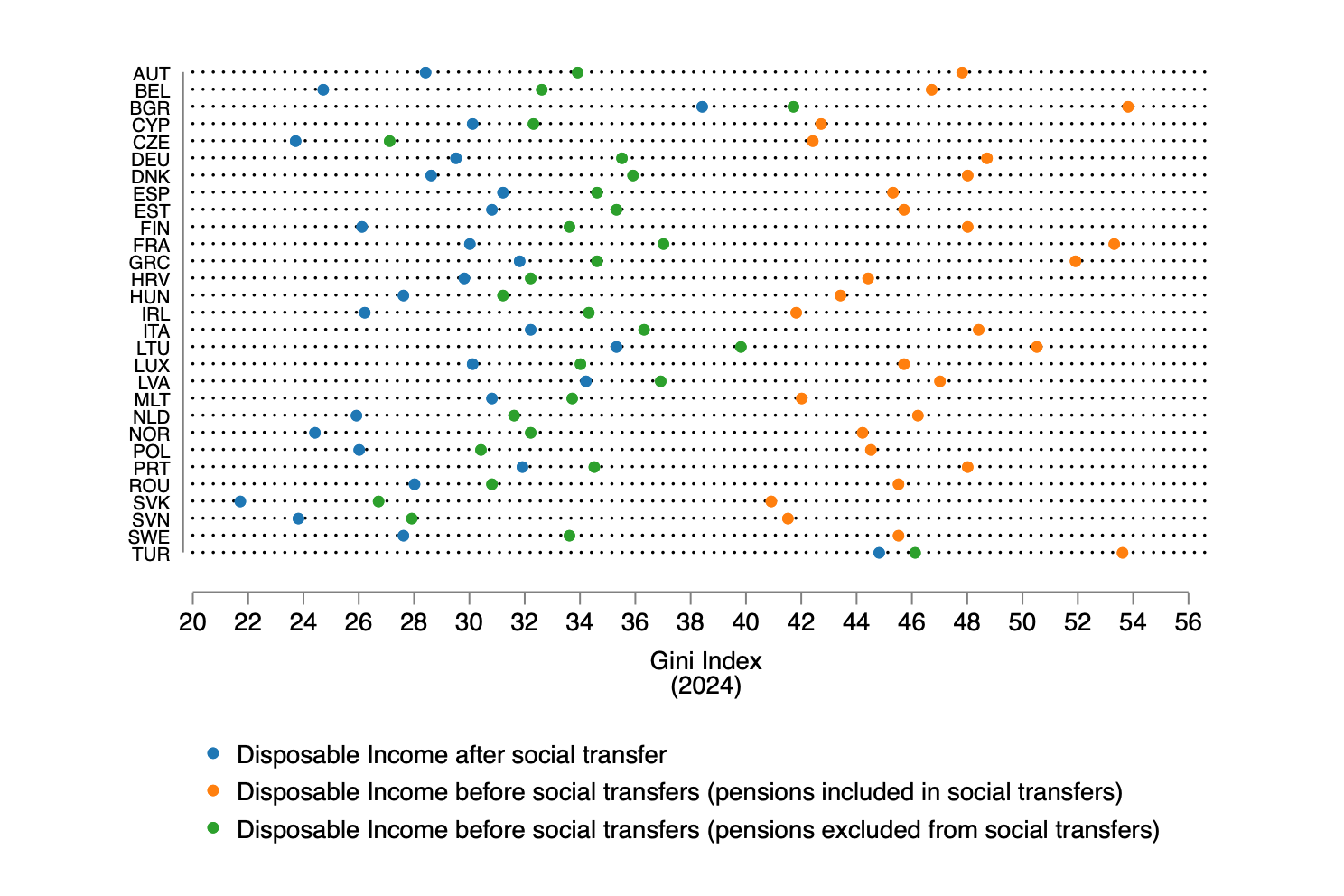}}
\caption{Ginis From Different Sub-metric Definitions (Eurostat, latest year)}
\label{fig:submetrics}
\end{figure}

Figure~\ref{fig:submetrics} illustrates this using data from Eurostat, which is distinctive in reporting Ginis for several explicitly defined variants of disposable income simultaneously. 

\medskip
\begin{center}\textit{[Figure 4 about here]}\end{center}
\medskip

The figure distinguishes Ginis computed from: (i)~equivalised disposable income; (ii)~equivalised disposable income before social transfers but including pensions; and (iii)~equivalised disposable income before all social transfers including pensions. The three distributions are clearly separated across countries, with pre-transfer Ginis substantially higher than post-transfer ones, as expected. The gaps between these variants are often large---measured in tens of percentage points---and are systematic across all European countries in the sample. Researchers who compare Ginis from different sources without checking whether transfers are included or excluded may be comparing fundamentally different economic concepts.

\subsection{Reference Unit}

The choice of reference unit---the unit over which welfare is measured and in whose name the distribution is computed---introduces additional comparability challenges. The main choices are between individual-level measures (with some form of adjustment for household size) and household-level measures (which treat the household as the unit of analysis). Within individual-level measures, the choice between per-capita income and income per adult equivalent (with a chosen equivalence scale) further affects distributional comparisons.

Considering the equivalence scale used in each country’s most recent database observation, a clear geographical pattern emerges: countries in the Global South predominantly report per capita measures, whereas countries in the Global North—particularly in Europe—tend to use adult-equivalent measures. For Latin American countries covered by SEDLAC, which reports inequality measures under both definitions, the difference between per capita and adult-equivalent Gini coefficients is typically only a few percentage points, rather than the larger gaps observed for welfare aggregates. Nevertheless, even differences of three to five percentage points can be consequential for cross-country comparisons, particularly where relatively small distributional differences carry significant policy implications.

\subsection{Bottom and Top Incomes}

Special attention is warranted for households with very low or very high reported incomes. Negative incomes can exist for legitimate accounting reasons---for example, farmers investing in one season and reaping returns in another---but it is unclear whether these households should be treated as poor (\citealp{HlasnyCerianiVerme_2022}). Zero incomes pose a smaller problem; our analysis confirms that including or excluding households with zero gross income does not substantially change the estimated Gini, consistent with these cases being a small share of all observations. To our knowledge, the World Bank’s Poverty and Inequality Platform (PIP) is the only database that explicitly documents the treatment of negative and zero incomes. In particular, all poverty and inequality indicators in PIP are computed using income and consumption distributions that exclude negative values and replace all observations below \$0.28 per person per day with \$0.28 per person per day (\citet{PIP_2024}, \S2.6).

Top incomes are more consequential. A well-documented stylised fact is that the very rich are systematically underrepresented in household surveys, either because they do not respond to surveys, respond with incomplete information, or are entirely excluded from sampling frames. A compounding issue is that statistical agencies routinely apply top-coding to microdata before public release---replacing the highest income values with a common cap or with synthetic values---primarily as a disclosure-control measure to protect the identity of high-income respondents. Top-coding therefore introduces a further downward distortion in measured inequality, over and above the bias already caused by non-response at the top. \citet{HlasnyVerme} show that for the United States, the Gini computed from public-access data (which are top-coded) and the Gini estimated after correcting for top-coding diverge by as much as 10 percentage points, with wide confidence intervals reflecting fundamental uncertainty about the upper tail of the income distribution.

\subsection{Post-Survey Adjustments}

Beyond the top-income problem, statistical agencies and database administrators routinely make several other types of post-survey adjustments. These include: spatial and inter-temporal price adjustments (using consumer price indices or Purchasing Power Parity conversion factors); imputations for missing or implausible values; top and bottom coding for data anonymisation; and, in some databases, large-scale re-calibration of survey distributions against National Accounts aggregates.

Each of these adjustments can affect measured inequality. The anchoring of survey distributions to National Accounts is especially significant: it is a central feature of the World Inequality Database's Distributional National Accounts (DINA) methodology \citep{BlanchetEtAl2024}, and it explains much of why WID Ginis---which incorporate top-income adjustments based on tax records and other administrative sources---tend to be considerably higher than survey-based alternatives (the mean Gini in WID is 55.4 versus 37.1 for the World Bank PIP; see Table~\ref{table:sumstats}).

\subsection{Comparability Across Time}

A dimension that receives less attention in discussions of cross-country comparability, but is no less important for empirical research, is comparability across time within a single country. Long time series of Gini coefficients---of the kind that appear in the WIID, SWIID, and WID for many countries---are susceptible to spurious trends arising from changes in survey design, the measurement of welfare aggregates, or the treatment of extreme observations. If a national statistical agency shifts from annual to monthly income measurement, revises its imputation procedure for missing income values, or alters the geographic coverage of its survey, the resulting Gini series will exhibit a structural break that can be mistaken for a genuine distributional trend.

The WID's DINA methodology introduces a further complication: the administrative records used to supplement surveys---primarily tax filings---have themselves evolved over time, as tax systems have been reformed, enforcement has tightened or relaxed, and reporting requirements have changed. A rising trend in WID Ginis for a given country could therefore reflect genuine concentration of income at the top, improved capture of top incomes in administrative records, or changes in tax law that alter what is reported. Distinguishing among these explanations requires access to the underlying administrative data and a detailed knowledge of each country's tax history---information that most users of global databases do not possess and that is not provided in database documentation.

For the databases in our collection, the periods with the densest coverage---roughly 2003 to 2017---also exhibit the smallest cross-database Gini ranges, in part because all databases draw on the same generation of national household surveys, and in part because metadata documentation has improved. Earlier decades show greater dispersion, partly because fewer databases cover those years, but also because the surveys underlying the estimates are more heterogeneous in design and geographic scope. Researchers relying on long time series from secondary databases should treat structural breaks in coverage---the entry or exit of a database, a change in the primary survey underlying the series, or a revision of historical estimates---as potential sources of spurious apparent trends.

\subsection{The Cumulative Challenge}

The six sources of discrepancy discussed above do not operate independently. Survey design choices determine which populations are included; survey implementation affects how accurately their welfare is measured; the welfare metric determines whether we measure income or consumption; sub-metric definitions determine which components are included in either aggregate; the reference unit determines how welfare is distributed across individuals within the household; and post-survey adjustments modify the raw distribution in ways that are often only partially disclosed.

Each choice introduces a layer of non-comparability, and their effects accumulate. A comparison between a gross income per-capita Gini from one country---typical of the Global North---and a consumption per-capita Gini from another---typical of the Global South---conflates at minimum two major sources of discrepancy before any differences in survey design or implementation are taken into account. The welfare metric alone accounts for a mean gap of approximately 5 Gini points globally, rising to 9 points for low-income countries; income sub-concept differences (gross versus net, with or without transfers) add several further points; and equivalence scale conventions contribute a smaller but non-negligible amount.

This cumulative challenge is what makes the study of global inequality so demanding and the interpretation of published Gini figures so hazardous for researchers who do not look behind the headline number. The twelve databases we review are not twelve independent measurements of the same underlying distribution; they are twelve different operationalisations of an inherently complex concept, each embedding methodological choices that are often undocumented, inconsistently applied across sources, and correlated with observable features of countries such as income level and region. Progress toward \emph{genuine} global comparability will therefore require agreement on welfare concepts, reference units, and sub-metric definitions at the level of primary data collection---not merely at the level of secondary database assembly.

\section{Quantifying Cross-Database Discordance}
\label{sec:quantifying}

Having catalogued the sources of discrepancy, we now quantify how much databases actually disagree in practice. We present four complementary exercises.

\subsection{Within-Country-Year Gini Variability}

For each country-year combination with observations from two or more databases, we compute the absolute range (maximum minus minimum Gini) and the standard deviation across database observations. Table~\ref{tab:variability_region} and Table~\ref{tab:variability_income} summarise these statistics by World Bank region and income group, respectively.

\medskip
\begin{center}\textit{[Table 3 about here]}\end{center}
\begin{center}\textit{[Table 4 about here]}\end{center}
\medskip

The results reveal striking heterogeneity. The mean within-country-year range across all 3,419 country-year pairs with at least two observations is 6.6 Gini points, with a median of 4.2 points. The highest variability is found in North America (mean range 9.1 pp) and Europe and Central Asia (8.9 pp), reflecting the particularly diverse income concepts used across databases for high-income countries. Sub-Saharan Africa and South Asia exhibit much lower variability (mean ranges of 3.6 and 3.6 pp respectively), consistent with these regions relying more uniformly on consumption-based welfare aggregates from a limited set of primary sources.

By income group, variability is highest for high-income countries (mean range 9.4 pp) and lowest for low-income countries (mean range 2.5 pp). This counterintuitive pattern arises because high-income countries have extensive coverage across multiple databases using different welfare concepts (income and consumption), while low-income countries are covered primarily by the global secondary databases, which largely draw on the same underlying survey sources. Maximum values reach 38.5 Gini points for some country-year pairs in Europe and Central Asia---a reminder that no two databases should be treated as interchangeable for individual country observations.

\subsection{Pairwise Cross-Database Concordance}

We compute, for each pair of databases sharing at least twenty overlapping country-year observations, both the Pearson correlation coefficient and the mean absolute difference (MAD) of their respective Gini values. Table~\ref{tab:concordance}  presents the full concordance matrix.

\medskip
\begin{center}\textit{[Table 5 about here]}\end{center}
\medskip

The concordance matrix reveals a clear structure. Databases that draw on the same primary microdata sources are highly correlated and show small MADs. The closest pair in Panel B is SEDLAC and WB-PIP, with a MAD of only 0.11 Gini points---reflecting that both rely heavily on the same Latin American household surveys. Similarly, ATG and SEDLAC (MAD = 0.66) and LIS and OECD (MAD = 0.88) are very close, reflecting shared primary sources and similar welfare concept coverage.

Eurostat is the clear outlier: its MAD with respect to all other databases ranges from 8.9 to 11.4 Gini points. This is explained by the fact that Eurostat covers only European countries, which have distinctively low inequality levels compared to the global sample, and uses a specific equivalised income concept (the OECD-modified scale) that is not replicated in databases with broader global coverage.

Notably, even among databases with high Pearson correlations---LIS and OECD have a correlation of 0.963---the MAD can still be substantial (0.88 pp on average across country-years). Correlation measures rank ordering; it does not preclude level differences that matter for cross-sectional inequality comparisons.

\subsection{The Income--Consumption Gap by Region}

Building on the bivariate comparisons in Figure~\ref{fig:incgross}, Tables~\ref{tab:inc_cons_region} and \ref{tab:inc_cons_income} document the income--consumption Gini gap across all country-year pairs in our unified database where both an income estimate and a consumption estimate are available for the same observation. The gap is positive and substantial across all world regions and income groups: income-based Ginis are consistently higher than consumption-based Ginis for the same country and year.

Regionally, the gap is widest in North America (mean 10.0 Gini points) and Sub-Saharan Africa (8.4 points), and narrowest in Europe and Central Asia (2.4 points). This pattern partly reflects the specific income sub-concepts prevalent in each region---gross income is common in North America while European databases often use net disposable income, reducing the gap relative to consumption---and partly reflects genuine differences in how income and consumption diverge as welfare aggregates across development contexts. By income group, the gap is largest for low-income countries (mean 9.0 Gini points) and declines to 3.1 points for high-income countries. This gradient is itself a reflection of the structural divide in welfare measurement practice documented in Section~\ref{sec:discrepancy}: low-income countries predominantly use consumption-based surveys while high-income countries use income-based ones. Any cross-country comparison that mixes the two metrics without adjustment will therefore tend to understate inequality in lower-income countries relative to higher-income countries by a margin that, at the extremes, approaches 9 Gini points---far larger than most genuine cross-country distributional differences of policy interest.

\medskip
\begin{center}\textit{[Table 6 about here]}\end{center}
\begin{center}\textit{[Table 7 about here]}\end{center}
\medskip

\subsection{Welfare Metric and Equivalence Scale: What the Evidence Shows}

Two further sources of descriptive evidence allow us to bound the contribution of methodology to measured inequality levels. First, the Eurostat data---unique in reporting three income concept variants for the same countries and years simultaneously---show that social transfers reduce the Gini by 10 to 20 percentage points in European countries (Figure~\ref{fig:submetrics}). A researcher who uses a pre-transfer Gini from one database and a post-transfer Gini from another for the same country will therefore dramatically overstate differences in underlying distributional structure. This is not a hypothetical risk: the global secondary databases (WIID, SWIID) pool observations from sources with different transfer treatment, and database metadata rarely specifies which income concept is in use for each country-year.

Second, the SEDLAC data confirm that shifting from per-capita to adult-equivalent income raises the Gini by roughly 2 to 4 points for Latin American countries. This is modest compared to the income--consumption gap, but not negligible for comparisons between countries at similar income levels where distributional differences are narrow.

These patterns indicate that methodological choices collectively drive much of the cross-database dispersion documented above. The upshot for users of global inequality data is direct: the Gini coefficient does not measure a single, well-defined quantity. The number reported by any given database is the product of a chain of methodological decisions---decisions that are often undocumented, inconsistently applied across sources, and correlated with observable features of countries such as income level and geographic region.

\section{A Country Example: Colombia}
\label{sec:casestudy}

The combination of the factors discussed above can produce dramatically different Gini values for the same country and year. To illustrate this point, we focus on Colombia, one of the countries with the highest number of data points in our aggregated database. The Gini series for Colombia covers the period 1964--2023 and, for many years, includes multiple observations from different databases using different welfare concepts, reference units, and methodological choices.

\begin{figure}[H]
{\centering \includegraphics[width=1\linewidth,height=\textheight,keepaspectratio]{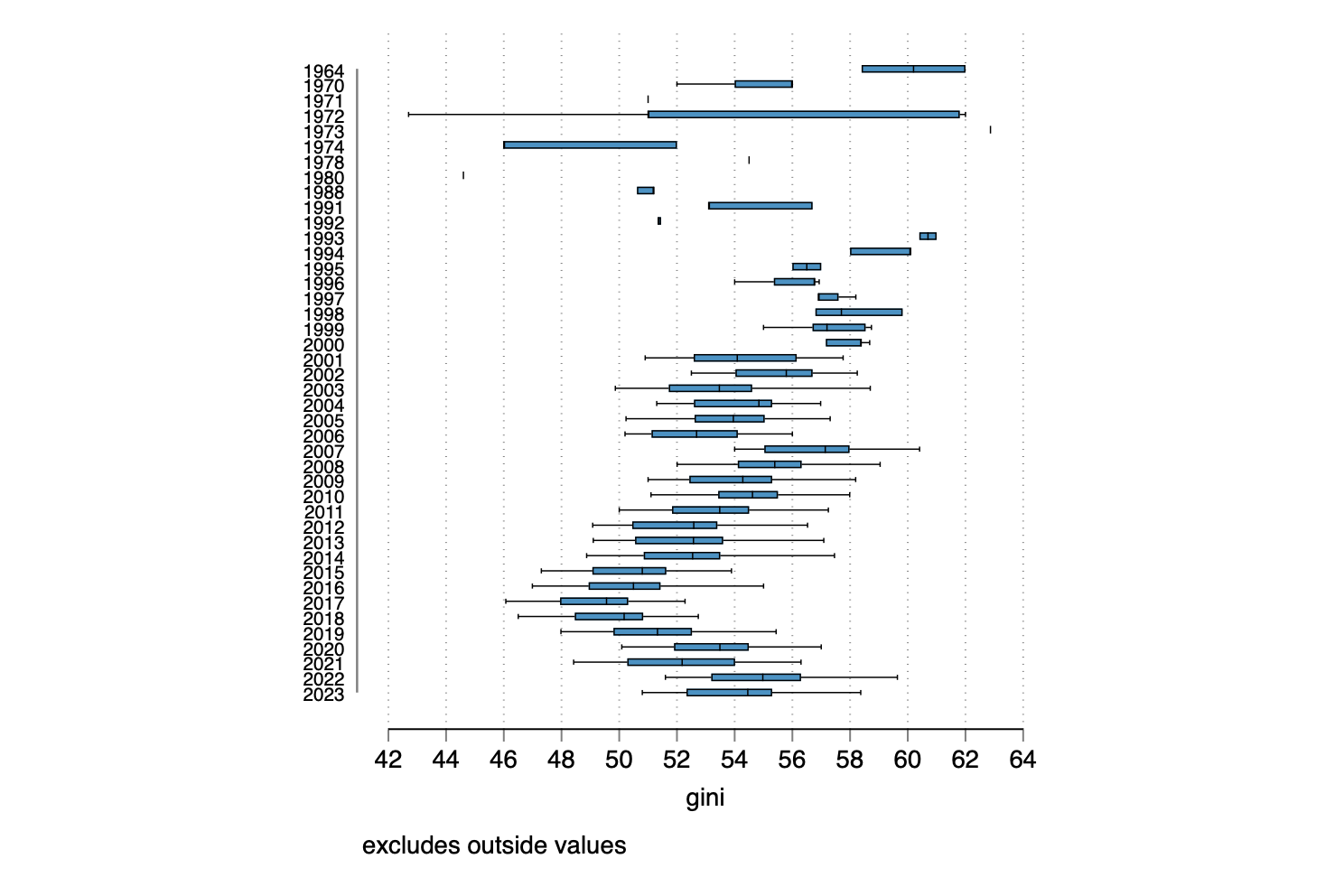}}
\caption{Distribution of Gini Estimates for Colombia by Year (all databases)}
\label{fig:colombia}
\end{figure}

Figure~\ref{fig:colombia} shows box plots of Gini estimates for Colombia by year.

\medskip
\begin{center}\textit{[Figure 6 about here]}\end{center}
\medskip

 The dispersion is substantial---gaps between minimum and maximum values often exceed 10 percentage points in years with at least four observations. The mean and median Gini also vary considerably across years, partly reflecting genuine distributional change and partly reflecting the changing composition of available estimates (which series are reported in which year varies across databases).

Colombia provides an especially instructive case because it is one of the most extensively covered countries in the databases focused on Latin America (IDB, SEDLAC, CEPAL, ATG) and is also included in global secondary databases (WIID, SWIID). The estimates span a wide range of welfare concepts---household per-capita income from surveys with different coverage of urban and rural areas, different treatments of zero incomes, and different periods of reference. A researcher interested in tracking Colombian inequality over the past 30 years faces a difficult task of selecting among these estimates, and the choice made can materially affect conclusions about the trend.

Authors' analysis on other countries confirms that the challenges illustrated for Colombia are general. For instance, Brazil, another high-coverage Latin American country, shows dispersion patterns similar to Colombia. India is covered almost exclusively by consumption-based measures, which limits cross-database dispersion but also makes comparison with income-based measures from other databases problematic. South Africa, despite having a larger share of income-based estimates than most Sub-Saharan African countries, shows wide dispersion reflecting the diversity of sources and welfare concepts. Germany, as a high-income country covered by many databases, shows the large income--versus--disposable-income gaps typical of the European context (Figures available upon request).

\section{Conclusion}
\label{sec:conclusion}

Professor Cornia foresaw the importance of monitoring inequality worldwide and pioneered some of the early efforts to assemble cross-country Gini data. His role in building and expanding the WIID at UNU-WIDER remains one of the most important institutional contributions to the scientific study of global inequality. The database he helped create has been the starting point for hundreds of studies examining the determinants and consequences of inequality across countries and over time. And his own use of these data---to document the inequality turn of the 1980s and 1990s, to track the rise and fall of inequality in Latin America, and to situate the distributional consequences of liberalisation and structural adjustment---set a standard for empirical rigour and policy relevance that continues to inspire.

This paper has shown that comparing Ginis across time and space remains technically demanding despite considerable progress over the past three decades. The twelve databases we have reviewed differ markedly in welfare concept, reference unit, sub-metric definitions, post-survey adjustments, and geographic and temporal coverage. These differences translate into large and systematic discrepancies in measured inequality. The income--consumption gap alone---averaging around 5 Gini points globally and exceeding 8--10 points in North America, South Asia, and Sub-Saharan Africa---is large enough to fundamentally affect cross-country comparisons and time-series analyses that mix the two welfare concepts without correction.

Our quantitative exercises add precision to this picture. Descriptive evidence from databases that simultaneously report multiple welfare concept variants---Eurostat for pre- and post-transfer income, SEDLAC for per-capita and adult-equivalent measures---confirms that sub-metric definitions and equivalence scale choices each introduce differences of several Gini points. The companion journal article presents a formal meta-regression that quantifies these effects, showing that welfare concept and equivalence scale together account for close to 80 percent of within-country-year cross-database variation. The pairwise concordance matrix confirms that databases sharing the same primary data sources are far more consistent than databases that aggregate across different welfare concepts. And the within-country-year variability analysis demonstrates that high-income countries, paradoxically, exhibit the highest cross-database dispersion---because they are covered by more databases employing diverse methodological approaches.

What would Professor Cornia have made of this situation? We think he would have found it concerning but not discouraging. Concerning, because the proliferation of databases without adequate metadata standards and without explicit harmonisation of welfare concepts creates real risks for empirical research: systematic differences across databases can introduce severe measurement error into growth, poverty, or political-economy regressions. Researchers who fail to account for these differences risk drawing conclusions that are artefacts of database choices rather than actual economic findings. Not discouraging, because the proliferation also reflects a vibrant research community and a growing institutional commitment to inequality measurement---a field to which Cornia contributed enormously.

The road ahead requires greater transparency in metadata, better harmonisation of welfare definitions across primary data collection efforts, and more systematic attention to the sources of cross-database discordance. This is what, we believe, Cornia would have wanted. And it is the best way we can honour the remarkable legacy he left us.


\bibliographystyle{plainnat}
\bibliography{References}

\newpage
\section*{List of Tables}

\begin{table}[H]
\centering
\begin{center}
\begin{footnotesize}
\begin{adjustbox}{angle=90}
\begin{tabular}{lcrclcp{1.3cm}ccl}
\toprule
 &&   & & & Wellbeing &  &  &  \\
Dataset & Primary & \# Obs & Countries & Time Range & Concept & Level & Sharing & Reference & Source\\
 & Dataset & & & & (\% Cons.)  & Unit & Unit &  \\
\midrule
\textbf{GLOBAL} &&&&&&&&\\
\midrule
ATG  & no & 5,120 & 200 &1948--2017 & 37 & overall & HH & Mixed$^{\dagger}$ &\citet{Milanovic_2019}\\
\hdashline
LIS & yes & 932 & 52 & 1963--2023& 0 &  & HH & AE &\citet{LIS_2024} \\
\hdashline
OECD IDD & no & 616 & 45 & 1974--2023 & 0 & overall   & HH & AE &\citet{OECD_2024} \\
\hdashline
UNU-WIDER WIID & no &  26,161 & 201 & 1867--2023 & 12   & overall  & HH & Mixed$^{*}$ &\citet{UNUWIDER_2025}\\
\hdashline
WB PIP & yes & 2,456 & 172 & 1963--2024&  40   & overall  & HH & PC &\citet{PIP_2024}\\
\hdashline
WID & yes & 10,419 & 217  & 1900--2023 & 0 & overall & Tax Unit& PC& \citet{BlanchetEtAl2024}\\
\hdashline
SWIID & no & 26,900 & 199  &  1960--2023&  14 &  overall &  \begin{minipage}[t]{1.5cm} \centering Ind \\ HH \end{minipage}  & Mixed$^{**}$ & \citet{Solt_2020}\\
\midrule
\multicolumn{9}{l}{\textbf{REGIONAL}}\\
\midrule
\multicolumn{9}{l}{\textbf{\textit{Africa}}}\\
Afristat & no & 84 & 20 & 1994--2023 & ? & overall  & ? & ? & \citet{Afristat_2025}\\
\midrule
\multicolumn{9}{l}{\textbf{\textit{East Asia and the Pacific}}}\\
ADB & no &322& 40 & 2000--2023  & 86 & overall & HH &  PC &\citet{ADB_2024, ADB_2025}\\
\midrule
\multicolumn{9}{l}{\textbf{\textit{Europe}}}\\
Eurostat & yes & 787 & 37 & 2014--2024 & 0 & \begin{minipage}[t]{1.5cm}  overall\\ $<$ 18 years \end{minipage} &HH  & AE & variable ilc\_di12c\\
\midrule
\multicolumn{9}{l}{\textbf{\textit{Latin America and the Caribbean}}}\\
CEPAL/ECLAC & yes & 959 & 19 & 2000--2023&  0 & \begin{minipage}[t]{1.5cm}  overall\\ urban/rural \end{minipage}& HH &  PC &\citet{CEPAL_2024}\\
\hdashline
IDB & yes & 45,656 & 26 & 1970--2023 & 0   & \begin{minipage}[t]{1.5cm} overall \\  urban/rural \\ gender \\migration  \\ ethnicity \\ disability  \\ quintiles \end{minipage} & \begin{minipage}[t]{0.5cm}  HH \\ Ind \end{minipage} & PC &\citet{IADB_2025}\\
\hdashline
SEDLAC & yes &442 & 24 & 1974--2023 & 0 & \begin{minipage}[t]{1.5cm} overall \\ urban \\ main cities \end{minipage} & HH &  PC/AE &\citet{SEDLAC_2024}\\
&&&&&&&&\\
\bottomrule
\multicolumn{9}{l}{$^*$ 39\% PC; 41\% AE; 20\% Household Based}\\
\multicolumn{9}{l}{$^\dagger$ 72\% Per capita; 8\% Household Based}\\
\multicolumn{9}{l}{$^{**}$ 32\% Per capita; 52\% AE; 15\% Household Based}\\
\end{tabular}
\end{adjustbox}
\end{footnotesize}
\end{center}
\captionof{table}{Overview of Databases Included in the Unified Collection\label{tab:databases}}
\end{table}

\begin{table}[H]
    \centering
    \caption{Summary Statistics by Dataset\label{table:sumstats}}
    \begin{footnotesize}
    \begin{tabular}{lrrccc}
    \toprule
    Dataset & \# Obs. & \# Countries & Min Year & Max Year & Mean Gini \\
    \midrule
    ADB         &     322 &  40 & 2000 & 2023 & 34.52 \\
    ATG         &   5,121 & 175 & 1948 & 2017 & 38.76 \\
    Afristat    &      84 &  20 & 1994 & 2023 & 39.40 \\
    CEPAL       &     887 &  18 & 2000 & 2023 & 47.10 \\
    Eurostat    &   2,102 &  37 & 2003 & 2024 & 37.71 \\
    IDB         &  45,656 &  26 & 1970 & 2023 & 36.69 \\
    LIS         &     930 &  52 & 1963 & 2023 & 33.70 \\
    OECD        &     616 &  45 & 1976 & 2023 & 31.85 \\
    SEDLAC      &     649 &  23 & 2000 & 2024 & 48.22 \\
    SWIID       &  26,900 & 199 & 1960 & 2023 & 39.49 \\
    UNU-WIDER   &  26,161 & 200 & 1867 & 2023 & 37.06 \\
    WB-PIP      &   2,504 & 172 & 1963 & 2024 & 37.17 \\
    WID         &  10,419 & 217 & 1900 & 2023 & 55.45 \\
    \midrule
    \textbf{Total} & \textbf{122,351} & \textbf{222} & \textbf{1867} & \textbf{2024} & \textbf{39.19} \\
    \bottomrule
    \end{tabular}
    \end{footnotesize}
\end{table}

\begin{table}[H]
    \centering
    \caption{Cross-Database Gini Variability by Region (country-years with $\geq 2$ database observations)\label{tab:variability_region}}
    \begin{footnotesize}
    \begin{tabular}{lrrrrr}
    \toprule
    Region & \# Obs & Mean Range & Mean SD & Median Range & Max Range \\
    \midrule
    East Asia \& Pacific (EAS) & 406 & 4.39 & 2.31 & 2.61 & 28.10 \\
    Europe \& Central Asia (ECS) & 1,532 & 8.94 & 3.89 & 6.52 & 38.51 \\
    Latin America \& Caribbean (LCN) & 631 & 5.15 & 2.26 & 4.27 & 45.25 \\
    Middle East \& N. Africa (MEA) & 200 & 3.18 & 1.62 & 1.91 & 19.66 \\
    North America (NAC) & 127 & 9.13 & 3.80 & 8.50 & 28.04 \\
    South Asia (SAS) & 182 & 3.57 & 1.90 & 2.15 & 22.52 \\
    Sub-Saharan Africa (SSF) & 341 & 3.63 & 2.01 & 1.92 & 27.65 \\
    \midrule
    \textbf{Total} & \textbf{3,419} & \textbf{6.55} & \textbf{2.97} & \textbf{4.20} & \textbf{45.25} \\
    \bottomrule
    \end{tabular}
    \end{footnotesize}
    \small\textit{\\Note: ``Range'' is maximum minus minimum Gini across databases for the same country-year. SD is standard deviation. All values in Gini points (pp).}
\end{table}

\begin{table}[H]
    \centering
    \caption{Cross-Database Gini Variability by Income Group (country-years with $\geq 2$ database observations)\label{tab:variability_income}}
    \begin{footnotesize}
    \begin{tabular}{lrrrrr}
    \toprule
    Income Group & \# Obs & Mean Range & Mean SD & Median Range & Max Range \\
    \midrule
    High income      & 1,334 & 9.42 & 4.14 & 7.30 & 38.51 \\
    Upper middle income & 811 & 4.99 & 2.25 & 3.63 & 31.72 \\
    Lower middle income & 484 & 3.33 & 1.69 & 2.15 & 26.52 \\
    Low income       & 126  & 2.46 & 1.34 & 1.07 & 24.08 \\
    \midrule
    \textbf{Total} & \textbf{2,755} & \textbf{6.73} & \textbf{3.03} & \textbf{4.41} & \textbf{38.51} \\
    \bottomrule
    \end{tabular}
    \end{footnotesize}
\end{table}

\begin{table}[htbp]
\centering
\caption{Pairwise Pearson Correlation and Mean Absolute Difference (pp) Across Databases}
\label{tab:concordance}

\rotatebox{90}{%
\begin{minipage}{\textheight} 
\centering

\scriptsize
\setlength{\tabcolsep}{3pt}
\renewcommand{\arraystretch}{0.9}

\begin{tabular}{lrrrrrrrrrr}
\toprule
 & ATG & CEPAL & Eurostat & IDB & LIS & OECD & SEDLAC & SWIID & UNU-WIDER & WB-PIP \\
\midrule
\multicolumn{11}{l}{\textit{Panel A: Pearson Correlation}} \\
ATG       & 1.000 & 0.915 & 0.327 & 0.817 & 0.944 & 0.919 & 0.972 & 0.783 & 0.870 & 0.958 \\
CEPAL     & 0.915 & 1.000 & .     & 0.830 & 0.862 & 0.492 & 0.964 & 0.861 & 0.889 & 0.922 \\
Eurostat  & 0.327 & .     & 1.000 & .     & 0.219 & 0.357 & .     & 0.095 & 0.234 & 0.332 \\
IDB       & 0.817 & 0.830 & .     & 1.000 & 0.855 & 0.550 & 0.834 & 0.580 & 0.649 & 0.791 \\
LIS       & 0.944 & 0.862 & 0.219 & 0.855 & 1.000 & 0.963 & 0.823 & 0.644 & 0.861 & 0.955 \\
OECD      & 0.919 & 0.492 & 0.357 & 0.550 & 0.963 & 1.000 & 0.414 & 0.462 & 0.684 & 0.917 \\
SEDLAC    & 0.972 & 0.964 & .     & 0.834 & 0.823 & 0.414 & 1.000 & 0.879 & 0.909 & 0.987 \\
SWIID     & 0.783 & 0.861 & 0.095 & 0.580 & 0.644 & 0.462 & 0.879 & 1.000 & 0.721 & 0.751 \\
UNU-WIDER & 0.870 & 0.889 & 0.234 & 0.649 & 0.861 & 0.684 & 0.909 & 0.721 & 1.000 & 0.886 \\
WB-PIP    & 0.958 & 0.922 & 0.332 & 0.791 & 0.955 & 0.917 & 0.987 & 0.751 & 0.886 & 1.000 \\
\midrule
\multicolumn{11}{l}{\textit{Panel B: Mean Absolute Difference (Gini points)}} \\
ATG       & 0.00 & 1.46 & 10.55 & 2.94 & 3.36 & 2.10 & 0.66 & 3.76 & 2.88 & 1.17 \\
CEPAL     & 1.46 & 0.00 & .     & 2.85 & 3.34 & 2.77 & 1.20 & 2.08 & 1.72 & 1.39 \\
Eurostat  & 10.55 & .    & 0.00  & .    & 11.42 & 10.27 & .    & 9.81 & 8.89 & 9.48 \\
IDB       & 2.94 & 2.85 & .     & 0.00 & 5.55 & 5.69 & 3.22 & 4.92 & 4.53 & 3.38 \\
LIS       & 3.36 & 3.34 & 11.42 & 5.55 & 0.00 & 0.88 & 3.20 & 6.73 & 3.85 & 2.64 \\
OECD      & 2.10 & 2.77 & 10.27 & 5.69 & 0.88 & 0.00 & 2.10 & 6.49 & 4.12 & 1.88 \\
SEDLAC    & 0.66 & 1.20 & .     & 3.22 & 3.20 & 2.10 & 0.00 & 1.88 & 1.40 & 0.11 \\
SWIID     & 3.76 & 2.08 & 9.81  & 4.92 & 6.73 & 6.49 & 1.88 & 0.00 & 4.26 & 3.71 \\
UNU-WIDER & 2.88 & 1.72 & 8.89  & 4.53 & 3.85 & 4.12 & 1.40 & 4.26 & 0.00 & 2.44 \\
WB-PIP    & 1.17 & 1.39 & 9.48  & 3.38 & 2.64 & 1.88 & 0.11 & 3.71 & 2.44 & 0.00 \\
\bottomrule
\end{tabular}

\medskip
\textit{Note: Pairs with fewer than 20 overlapping country-year observations are denoted by a period.}

\end{minipage}%
} 
\end{table}

\begin{table}[H]
    \centering
    \caption{Income--Consumption Gini Gap by Region\label{tab:inc_cons_region}}
    \begin{footnotesize}
    \begin{tabular}{lrrrr}
    \toprule
    Region & \# Obs & Mean Gap & Median Gap & P75 Gap \\
    \midrule
    East Asia \& Pacific      & 133 & 4.52 & 4.18 & 7.49 \\
    Europe \& Central Asia    & 408 & 2.43 & 2.18 & 5.42 \\
    Latin America \& Caribbean& 98  & 8.50 & 8.12 & 11.78 \\
    Middle East \& N. Africa  & 78  & 3.59 & 3.46 & 5.40 \\
    North America             & 46  & 10.04& 11.30& 12.75 \\
    South Asia                & 61  & 7.49 & 6.23 & 10.15 \\
    Sub-Saharan Africa        & 63  & 8.44 & 7.22 & 13.80 \\
    \midrule
    \textbf{Total}            & 887 & 4.68 & 4.37 & 7.46 \\
    \bottomrule
   \multicolumn{5}{p{11cm}}{\textit{Note: Gap defined as income Gini minus consumption Gini for same country-year. All values in Gini points.}
}
    \end{tabular}
    \end{footnotesize}\\
\end{table}

\begin{table}[H]
    \centering
    \caption{Income--Consumption Gini Gap by Income Group\label{tab:inc_cons_income}}
    \begin{footnotesize}
    \begin{tabular}{lrrr}
    \toprule
    Income Group & \# Obs & Mean Gap & Median Gap \\
    \midrule
    High income          & 367 & 3.06 & 2.49 \\
    Upper middle income  & 295 & 5.46 & 5.13 \\
    Lower middle income  & 157 & 6.78 & 5.63 \\
    Low income           &  28 & 9.00 & 7.83 \\
    \midrule
    \textbf{Total}       & 847 & 4.78 & 4.42 \\
    \bottomrule
    \end{tabular}
    \end{footnotesize}
\end{table}

\newpage
\section*{List of Figures}

\end{document}